\documentclass[aps,pra,eqsecnum,twocolumn]{revtex4-1}
\usepackage{amssymb}
\usepackage{amsmath}
\usepackage{xcolor}
\definecolor{maroon}{RGB}{150,0,0}
\definecolor{dblue}{RGB}{0,0,200}
\usepackage[colorlinks=true,linkcolor=dblue,citecolor=maroon,urlcolor=maroon]{hyperref}
\begin{document}
\title{Properties of Spin and Orbital Angular Momenta of Light}
\author{Arvind}
\email{arvind@iisermohali.ac.in}
\affiliation{Department of Physical Sciences, Indian
Institute of Science Education and Research (IISER) Mohali,
Sector 81 SAS Nagar, Manauli PO 140306, 
Punjab, India}
\author{S. Chaturvedi}
\email{subhash@iiserb.ac.in}
\affiliation{Department of Physics, Indian Institute of
Science Education and Research (IISER) Bhopal, Bhopal Bypass
Road, Bhauri, Bhopal 462066, India}
\author{N. Mukunda}
\email{nmukunda@gmail.com}
\affiliation{Adjunct Professor, Department of Physics,
Indian Institute of Science Education and Research (IISER)
Bhopal, Bhopal Bypass Road, Bhauri, 
Bhopal 462066, India }
\begin{abstract}
This paper analyses the algebraic and physical properties of
the spin and orbital angular momenta of light in the quantum
mechanical framework. The consequences of the fact that
these are not angular momenta in the quantum mechanical
sense are worked out in mathematical detail.  It turns
out that the spin part of the angular momentum has continous
eigen values. Particular attention is given to the paraxial
limit, and to the definition of Laguerre – Gaussian modes
for photons as well as classical light fields taking full
account of the polarization degree of freedom.
\end{abstract}
\maketitle
\section{Introduction}
\label{introduction}
There has been great interest for some time now in the
angular momentum properties of the Maxwell
field~\cite{babiker-book}, in particular its proposed
separation into what have been called spin and orbital
angular momentum of light~\cite{wolf-book}. In addition to many theoretical
investigations~\cite{allen_book,
enk-epl-94, barnett_JMO, Bialynicki_Birula_2011,
Bliokh_2014, Bialynicki_Birula_2014, barnett-ptrsla-17,
calvo-pra-06,apt1,apt2,apt3,apt4,apt5,apt6,apt7,apt8,apt9,apt10,deutsch-pra-91} extensive experimental
work~\cite{ape2,ape3,ape4,ape9,orbital_spin,jha,twisted_photons} has also been devoted to
understand these concepts.

In an earlier work~\cite{oam1} a unified framework for studying these
novel properties of light, in both classical and quantum
domains, has been presented. In particular, the fact that
the spin and orbital parts of the total angular momentum are
not truly quantum mechanical angular momenta at all has
been emphasized.

The aim of the present paper is to carry this study further
and in particular to analyse in full detail the quantum
mechanical properties of the spin angular momentum of light
at the one photon level. It is seen that the eigenvalues and
eigenvectors of the spin angular momentum are very different
from those of any true angular momentum as understood in
quantum mechanics. The essential roles of polarization and
transversality of light are brought out, and new vectorial
Laguerre--Gauss fields including polarization in the
paraxial regime are constructed.

The contents of this paper are organised as follows.
Section~\ref{background}
reviews the formulation of the free Maxwell  equations in
a particularly economical form  using the complex transverse
analytic signal vector potential. The seven basic constants
of motion following from Poincar\'e invariance are expressed
in terms of analytic signal vector potential and electric
field. The spin and orbital angular momenta, SAM and OAM,
which are also real constants of motion,  are identified.
The description of a general solution of the Maxwell
equations using a complex transverse vector function on wave
vector space, and a natural Lorentz invariant Hilbert space
made up of such functions, is outlined. Canonical
quantisation is recalled, and the operator forms of the
seven hermitian constants of motion, as well as of the SAM
and OAM, are listed. A convenient description of the set of
all single photon states in terms of the classical Hilbert
space is mentioned.  The rest of this paper deals
essentially with one photon states. In
Section~\ref{spin_orbital} some of the
properties of the SAM and OAM  operators are worked  out.
The connections  to the helicity operator,  the component
of the total angular momentum in the momentum direction, are
obtained and its properties are described. Helicity is a
well defined concept in terms of the generators of the
Poincar\'e group. The fact that the SAM components are
commutative, and that along with the total angular momentum
they generate  a Euclidean group, is brought out. The
helicity operator is seen to be invariant under this
Euclidean group. For later comparison, the discrete set of
complete orthonormal  eigenfunctions of total angular
momentum are recorded.  Section~\ref{helicity} solves completely the
problem of eigenvalues and eigenfunctions for the SAM along
with helicity. It is emphasised that these are ideal non
normalisable eigenfunctions, as the eigenvalues of the SAM
components are continuous. The contrast with the total
angular momentum eigenfunctions is explicitly seen. To
emphasize this aspect, the properties of SAM in a normalised
simultaneous eigenfunction of the helicity and the third
component of total angular momentum are worked out. It is
shown that such an eigenfunction can never be an
eigenfunction of the third component of SAM as well; the
SAM components have a nontrivial variance matrix in such a
state.   Section~\ref{paraxial} is devoted to an analysis of the paraxial
regime. It is recalled that it is appropriate to perform
canonical quantisation before considering the paraxial
limit. The approximate nature of this limit, and the
correspondingly approximate consequence  of transversality,
are both clearly brought out. These considerations,
combined with the paraxial limit of the general simultaneous
eigen functions of helicity and third component of total
angular momentum, lead to the development of Laguerre –
Gauss mode functions for the vector Maxwell field.  The
helicity eigenvalue of plus or minus $\hbar$ appears as a
third label added to the two that  enumerate  the modes in
the scalar optical  case.  Section~\ref{conc} is devoted to
Concluding Remarks.

\section{Constants of motion and quantization of the free
Maxwell field}
\label{background}
We begin with the classical free Maxwell equations written
in terms of the complex positive frequency analytic signal
vector potential 
${\bf A}^{(+)}(x)$ where $x\equiv ({\bf x},t)$. The basic
(first order) equation of motion (EOM) is 
\begin{eqnarray}
i\dfrac{\partial}{\partial t}{\bf A}^{(+)}({\bf x},t)
&=&(\hat{\omega}{\bf A}^{(+)})({\bf x},t), \nonumber \\
 \hat{\omega}&=& c (-\mbox{\boldmath$\nabla$}^2)^{1/2}.
\label{2.1}
 \end{eqnarray}
This is consistent with the transversality constraint  
\begin{eqnarray}
\mbox{\bf{$\nabla$}}\cdot{\bf A}^{(+)}({\bf x},t)=0.
\label{2.2}
\end{eqnarray}
The initial data is specified by ${\bf A}^{(+)}({\bf x},
0)$. The analytic signal electric and magnetic fields can be
regarded as 
derived quantities at each instant of time :
\begin{eqnarray}
{\bf E}^{(+)}(x)=\dfrac{i}{c}(\hat{\omega}{\bf A}^{(+)})(x),
~~~~{\bf B}^{(+)}(x)=\mbox{\boldmath$\nabla$}\wedge{\bf
A}^{(+)}(x).\nonumber\\
\end{eqnarray}
They are also transverse and obey first order EOM similar to
${\bf A}^{(+)}$ in~(\ref{2.1}). For convenience we will
use both 
${\bf A}^{(+)}$ and ${\bf E}^{(+)}$ in various important expressions. 

From the relativistic invariance of the Maxwell equations we
obtain seven constants of motion (COM) which have no
explicit time dependence --
momentum ${\bf P}$, energy $P^{0}$, and total angular
momentum ${\bf J}$ (all real):
\begin{eqnarray}
P_j&=&\dfrac{1}{2\pi c}\int d^3x ~{\bf E}^{(+)}(x)^\ast
\cdot \partial_j{\bf A}^{(+)}(x),\nonumber\\
 {P}^{0}&=&\dfrac{1}{2\pi}\int d^3x~
{\bf E}^{(+)}(x)^\ast\cdot \partial^0{\bf A}^{(+)}(x),
\nonumber \\
J_j&=&\dfrac{1}{2\pi c}\int d^3x ~{ E}_m^{(+)}(x)^\ast
\left(\delta_{mn}({\bf x}\wedge
\mbox{\boldmath$\nabla$})_j \right.
\left. +\epsilon_{jmn}\right){ A}_n^{(+)}(x).\nonumber\\
\label{2.4}
\end{eqnarray}
Here $(\partial^{0}\equiv \dfrac{\partial}{\partial x_0},
x_0=-x^0=-ct)$. The two terms in the total angular momentum
${\bf J}$ are identified as 
the orbital angular momentum (OAM) and spin angular momentum
(SAM) respectively of the free field, and both are real
COM's:
\begin{eqnarray}
L_j&=&\dfrac{1}{2\pi c}\int d^3x ~{ E}_m^{(+)}(x)^\ast ({\bf
x}\wedge \mbox{\boldmath$\nabla$})_j{A}_m^{(+)}(x)
,\nonumber\\
S_j&=&\dfrac{1}{2\pi c}\int d^3x\ \epsilon_{jmn}
~{E}_m^{(+)}(x)^\ast{A}_n^{(+)}(x).
\end{eqnarray}
These will be studied in detail in the sequel. 

The general solution of (\ref{2.1}) and (\ref{2.2}) can be
written in terms of a complex transverse function ${\bf
v}({\bf k})$ of the real wave 
vector ${\bf k}\in \mathbb{R}^3$ :
 \begin{eqnarray}
{\bf A}^{(+)}({\bf x},t)&&=\dfrac{c}{2\pi}
\int\dfrac{d^3k}{\sqrt{\omega}}e^{ik\cdot x}{\bf v}({\bf k}),\nonumber\\
{\bf E}^{(+)}({\bf x},t)&&=
\dfrac{i}{2\pi}\int d^3k~\sqrt{\omega}
e^{ik\cdot x}{\bf v}({\bf k}), \nonumber\\
{\bf k}\cdot {\bf v}({\bf k})=0,&& \,\,\omega=ck=c|{\bf k}|,~~
k\cdot x={\bf k}\cdot{\bf x}-\omega t.
\end{eqnarray}
Thus the most general free Maxwell field is given equally
well by ${\bf A}^{(+)}(x)$ or ${\bf v}({\bf k})$. The seven
COM's~(\ref{2.4}) can be 
expressed in terms of  ${\bf v}({\bf k})$ :
 \begin{eqnarray}
  P_j&=&\int d^3k ~k_j ~{\bf v}({\bf k})^\ast\cdot {\bf
v}({\bf k}), ~~ P^0=\int d^3k~ \omega ~{\bf v}({\bf
k})^\ast\cdot {\bf v}({\bf k})\nonumber\\
   J_j&=&\int d^3k  ~{ v}_m({\bf k})^\ast
(-i\delta_{mn}({\bf k}\wedge
\tilde{\mbox{\boldmath$\nabla$}})_j-i\epsilon_{jmn}) {
v}_n({\bf k}), 
   \nonumber\\
   \tilde{\partial}_j&=&\dfrac{\partial}{\partial k_j}.
 \end{eqnarray}
 The OAM and SAM are 
 \begin{eqnarray}
L_j&=&-i\int d^3k ~{ v}_m({\bf k})^\ast ({\bf
k}\wedge \tilde{\mbox{\boldmath$\nabla$}})_j{v}_m({\bf k})
,\nonumber\\
S_j&=&-i\int d^3k\ 
~{v}_m({\bf k})^\ast \epsilon_{jmn}{v}_n({\bf k}).
\end{eqnarray}
At the classical level we define a Hilbert space ${\cal M}$
by using a metric in the space of amplitudes ${\bf v}({\bf
k})$:
\begin{align}
{\cal M} =\big{\{}{\bf v}({\bf k}) ~\big{|}~{\bf k}\cdot
{\bf v}({\bf k})&=0, \nonumber\\
~~~~||{\bf v}||^2 = \int d^3k~&{\bf v}({\bf k})^\ast \cdot {\bf v}({\bf k}) < \infty\big{ \}}.
\label{2.9}
\end{align}
The norm $||{\bf v}||$ is Lorentz invariant. The space
${\cal M}$ will play an important role after quantization to
which we now turn.

The process of canonical quantization involves replacing the
classical amplitudes ${\bf v}({\bf k}), {\bf v}({\bf
k})^\ast$ by vectorial 
operators $\sqrt{\hbar}\hat{{\bf a}}({\bf k}),
\sqrt{\hbar}\hat{{\bf a}}({\bf k})^\dagger $ obeying the
canonical commutation relations 
(CCR) on a suitable Hilbert space ${\cal H}$:
\begin{eqnarray}
&& [\hat{a}_j({\bf k}), \hat{a}_l({\bf k}^\prime)^\dagger] =
\left(\delta_{jl}-\dfrac{k_jk_l}{|{\bf
k}|^2}\right)\delta^{(3)}({\bf k}-{\bf
k}^\prime),\nonumber\\&& [{\bf \hat{a}}, {\bf \hat{a}}]=
[{\bf \hat{a}}^\dagger, {\bf \hat{a}}^\dagger]=0,\nonumber\\
&&{\bf k}\cdot \hat{a}({\bf k})={\bf k}\cdot \hat{a}({\bf k})^\dagger=0.
\label{CCR}
\end{eqnarray}
The field operators are 
\begin{eqnarray}
&& \hat{{\bf A}}^{(+)}(x)=\dfrac{c}{2\pi}\sqrt{\hbar}\int
\dfrac{d^3k}{\sqrt{\omega}}~e^{ik\cdot x}\hat{{\bf a}}({\bf
k}),\nonumber\\
&& \hat{ {\bf E}}^{(+)}(x)=\dfrac{i}{2\pi}\sqrt{\hbar}\int
d^3k~ \sqrt{\omega}e^{ik\cdot x}\hat{{\bf a}}({\bf k}).
\end{eqnarray}
The operator forms of the classical COM's are the hermitian operators
\begin{eqnarray}
\hat{P}^{0}
&=&\int d^3k ~\hbar\omega~
\hat{{\bf a}}({\bf k})^\dagger \cdot \hat{{\bf a}}({\bf k});\nonumber\\
\hat{P}_j
&=& \int d^3k ~\hbar k_j ~\hat{{\bf a}}({\bf
k})^\dagger\cdot \hat{{\bf a}}({\bf k});\nonumber\\
\hat{J}_j
&=&-i\hbar \int d^3k ~\hat{a}_m({\bf k})^\dagger
(\delta_{mn} ({\bf k}\wedge
\tilde{\mbox{\boldmath$\nabla$})}_j+\epsilon_{jmn})
~\hat{a}_n({\bf k});~~(a)\nonumber\\
\hat{L}_j&=&-i\hbar\int d^3k~\hat{a}_m({\bf k})^\dagger({\bf
k}\wedge \tilde{\mbox{\boldmath$\nabla$}})_j\hat{a}_m({\bf
k}),\nonumber\\
\hat{S}_j&=&-i\hbar\int d^3k~\hat{a}_m({\bf k})^\dagger
\epsilon_{jmn}\hat{a}_n({\bf k}).~~~~~~~~~~~~~~~~~~~~~~~~~~(b)\nonumber\\
\end{eqnarray}
The commutation relations among the former are determined by
the Poincar\'e group structure:
\begin{eqnarray}
&&[\hat{P}^\mu, \hat{P}^\nu] = 0;\nonumber\\
&& [\hat{J}_j,\hat{P}^0]=0;~~ [\hat{J}_j,\hat{P}_l]
=i\hbar \epsilon_{jln}\hat{P}_n;\nonumber\\
&& [\hat{J}_j, \hat{J}_l]= i\hbar\epsilon_{jln}\hat{J}_n.
\label{2.13}
\end{eqnarray} 
We will examine the important operator properties of the OAM
and SAM, $\hat{L}_j$ and ${\hat S}_j$, in the next Section. 
 
The Hilbert space ${\cal H}$ on which the CCR's~(\ref{CCR})
 are realized irreducibly is the direct sum of
subspaces 
${\cal H}_n, n=0,1,2,\cdots,$ made up of states with
definite total photon number $n$. Thus ${\cal H}_0$ is the
one dimensional 
 subspace of no photon states ( multiples of the vacuum
state $|0\rangle$); ${\cal H}_1$ is the subspace of single
photon states; and so on. 
 The importance of the classical Hilbert state ${\cal M}$,
Eq.~(\ref{2.9}), is that there is a one to one
correspondence ${\cal M} 
 \leftrightarrow {\cal H}_1$, given by the following structure:
\begin{eqnarray}
&&{\bf v}({\bf k})\in {\cal M}, ~~~~~|{\bf v}\rangle={\hat
a}({\bf v})^\dagger|0\rangle \in {\cal H}_1, \nonumber\\
&& {\hat a}({\bf v})=\dfrac{1}{\sqrt{\hbar}}\int {d^3k}~
{\bf v}({\bf k})^\ast \cdot \hat{{\bf a}}({\bf k}), \nonumber\\
&&{\hat a}({\bf v})^\dagger
=\dfrac{1}{\sqrt{\hbar}}\int {d^3k}~ {\bf v}({\bf k}) \cdot
\hat{{\bf a}}({\bf k})^\dagger; \nonumber\\
&& [{\hat a}({\bf v}),{\hat a}({\bf
v}^\prime)^\dagger]=\dfrac{({\bf v},{\bf
v}^\prime)}{\hbar}\mathbb{I};\nonumber\\
&& \hat{a}_j({\bf k})|{\bf v}\rangle=\dfrac{1}{\sqrt{\hbar}}v_j({\bf k})|0\rangle.
\label{2.14}
\end{eqnarray}
The inner products among one photon states in ${\cal H}_1$
are essentially the classical inner products in ${\cal M}$: 
\begin{equation}
 \langle {\bf v}^\prime|{\bf v}\rangle = ({\bf v}^\prime, {\bf v})/\hbar
\end{equation}

\section{Operator properties of total, orbital and spin
angular momentum of photons}
\label{spin_orbital}
We now take up a detailed analysis of the operators
$\hat{{\bf L}}$, $\hat{{\bf S}}$ representing the OAM and
SAM of the quantized Maxwell field respectively. For 
our purposes it suffices to restrict these (and other)
operators to one-photon states in ${\cal H}_1$. Their
actions on a one-photon wavefunction
${\bf v}({\bf k})$ can be expressed in a succinct manner.
For $\hat{P}^\mu$ and $\hat{{\bf J}}$ we have:
\begin{eqnarray}
&&(\hat{P}^0|{\bf v}\rangle)_j({\bf k}) = \hbar\omega
v_j({\bf k})~, ~~(\hat{P}_l|{\bf v}\rangle)_j({\bf k}) =
\hbar k_l v_j({\bf k}),\nonumber\\
&&  (\hat{J}_l|{\bf v}\rangle)_j({\bf k})= -i\hbar \left(
({\bf k}\wedge \tilde{\mbox{\boldmath$\nabla$}})_l v_j({\bf
k})+ \epsilon_{ljn}
  v_n({\bf k})\right).
  \label{3.1}
  \end{eqnarray}
  For $\hat{{\bf L}}$ and $\hat{{\bf S}}$ we find:
 \begin{eqnarray}
(\hat{L}_l|{\bf v}\rangle)_j({\bf k})&=& -i\hbar \left(
({\bf k}\wedge \tilde{\mbox{\boldmath$\nabla$}})_l v_j({\bf
k})
    + \dfrac{k_j}{|{\bf k}|^2} ({\bf k}\wedge {\bf v}({\bf
k}))_l\right),\nonumber\\
    (\hat{S}_l|{\bf v}\rangle)_j({\bf k})&=&  i\hbar
\dfrac{k_l}{|{\bf k}|^2} ({\bf k}\wedge {\bf v}({\bf k}))_j. 
    \label{3.2}
    \end{eqnarray}
    Two operator relations follow easily :
 \begin{eqnarray}
  \hat{{\bf P}}\cdot \hat{{\bf L}}&=&0~~, ~~ \hat{{\bf
P}}\wedge \hat{{\bf S}}=0~~.
  \label{3.3}
 \end{eqnarray}
 The helicity operator $\hat{W}$ is defined in terms of
Poincar\'e group generators as 
 \begin{eqnarray}
  \hat{W}= \dfrac{\hat{{\bf P}}\cdot \hat{{\bf J}}}{\sqrt{
\hat{{\bf P}}\cdot \hat{{\bf P}}}}.
 \end{eqnarray}
 With~(\ref{3.3}) this simplifies to
 \begin{eqnarray}
  \hat{W}= \dfrac{\hat{{\bf P}}\cdot \hat{{\bf S}}}{\sqrt{
\hat{{\bf P}}\cdot \hat{{\bf P}}}}~.
 \end{eqnarray}
We next easily find some operator product relations:
   \begin{eqnarray}
\hat{{\bf J}}\cdot \hat{{\bf S}}= \hat{{\bf S}}\cdot
\hat{{\bf S}}=\hat{W}^2=\hbar^2.
  \label{3.6}
 \end{eqnarray}
 Therefore we also have
\begin{eqnarray}
  \hat{{\bf L}}\cdot \hat{{\bf S}} =0.
\end{eqnarray}
Turning to commutators, while Eqs.~(\ref{2.13}) are part of
the Poincar\'e Lie algebra, we now find these additional
ones:
\begin{eqnarray}
&&[\hat{J}_l, \hat{W}]=0 ;\nonumber\\
 &&[\hat{J}_l, \hat{L}_m\ \ \mbox{or}\ \ \hat{S}_m]=i\hbar\
\epsilon_{lmn}(\hat{L}_n\ \ \mbox{or}\
\hat{S}_n);\nonumber\\
 &&[\hat{S}_l, \hat{P}_m\ \ \mbox{or}\ \ \hat{S}_m \ \
\mbox{or}\ \ \hat{W}]=0~,~~ [\hat{L}_l, \hat{W}]=0.
\end{eqnarray}
 As expected, ${\hat W}$ is a rotational scalar while
$\hat{{\bf L}}$ and $\hat{{\bf S}}$ are vectors. The six
hermitian operators 
$\hat{{\bf J}}$ and $\hat{{\bf S}}$, all having the
dimensions of action, realise the Lie algebra of a Euclidean
group $\tilde{E}(3)$. 
This is distinct from the Euclidean subgroup $E(3)$ of the 
Poincar\'e group, generated by $\hat{{\bf J}}$ and $\hat{{\bf P}}$.

The result for ${\hat W}^2$ in Eq.~(\ref{3.6}) seems
counterintuitive, since all $\hat{P}_j$ and $\hat{S}_j$
commute pairwise and all have continuous eigenvalues. The
reason of course is the result $\hat{{\bf P}}\wedge
\hat{{\bf S}}=0$.

The operators $\hat{{\bf J}}$ constitute a quantum
mechanical angular momentum. Thus the eigenvalues of
$\hat{{\bf J}}\cdot \hat{{\bf J}}$ and $\hat{J}_3$ are
$l(l+1)\hbar^2$ and $\hbar m$ respectively, for
$l=1,2,\cdots,$ and $m=l,l-1,\cdots, -l$ for photons. As is
well known, their simultaneous eigenfunctions form a
complete orthonormal basis for transverse vector functions
of the unit wave vector $\hat{{\bf k}}\in
\mathbb{S}^2$~\cite{blatt-book,devaney-jmp-74}~:
\begin{eqnarray}
&&\{\hat{{\bf J}}^2,~\hat{{\bf J}}_3\} {\bf
Y}^{(a)}_{lm}(\hat{{\bf k}})=\{\hbar^2l(l+1),m\hbar\} {\bf
Y}^{(a)}_{lm}(\hat{{\bf k}}), 
 \quad a=1, 2;\nonumber\\ 
&&{\bf Y}^{(1)}_{lm}(\hat{{\bf
k}})=\dfrac{1}{\sqrt{l(l+1)}} (-i{\bf k}\wedge
\tilde{\mbox{\boldmath$\nabla$}}){Y}_{lm}(\hat{{\bf
k}}),\nonumber\\
&&{\bf Y}^{(2)}_{lm}(\hat{{\bf k}})=\hat{{\bf k}}\wedge
{\bf Y}^{(1)}_{lm}(\hat{{\bf k}});\nonumber\\
&&\int_{\mathbb{S}^2} d\Omega(\hat{{\bf k}}) {\bf
Y}^{(a^\prime)}_{l^\prime m^\prime}(\hat{{\bf k}})^\ast\cdot
{\bf Y}^{(a)}_{lm}(\hat{{\bf k}})
  =\delta_{a^\prime,a}\delta_{l^\prime,l}\delta_{m^\prime,m}
;\nonumber\\
&& \sum_{a=1}^{2}\sum_{l=1}^{\infty}\sum_{m=-l}^{l}
 {Y}^{(a)}_{l m,j}(\hat{{\bf k}}) {
Y}^{(a)}_{lm,j^\prime}(\hat{{\bf k}}^\prime)^\ast
 =\delta^{(2)}(\hat{{\bf k}},\hat{{\bf
k}}^\prime)(\delta_{jj^\prime}-
\dfrac{k_jk_{j^\prime}}{|{\bf k}|^2}).\nonumber\\
\label{3.9}
\end{eqnarray}
Here $Y_{lm}(\hat{{\bf k}})$ are the usual spherical
harmonics and $\delta^{(2)}(\hat{{\bf k}},\hat{{\bf
k}}^\prime)$ is the two dimensional surface Dirac delta
function over $\mathbb{S}^2$.

As we will see, since the $\hat{{\bf S}}$ are not an angular
momentum, their eigenvalues and eigenvectors have very
different characters. 
\section{Spin and helicity eigenfunctions, variance matrix
for spin}
\label{helicity}
Now we consider the eigenvalues and eigenvectors of the SAM
$\hat{{\bf S}}$. Since the four operators
$\hat{S}_j,\hat{W}$ commute pairwise, they can all be
simultaneously diagonalized. As the $\hat{S}_j$ transform as
a three dimensional vector under spatial rotations, we see
from Eqs.~(\ref{3.6}) that the possible eigenvalues for
$\hat{{\bf S}}$ and ${\hat W}$ have the forms
\begin{eqnarray}
\hat{{\bf S}}\rightarrow \hbar{\bf s},~~{\hat W}\rightarrow
\hbar w, ~~{\bf s} \in S^{2}, w=\pm 1.
 \end{eqnarray}
 It follows that while $\hat{W}$ possesses normalizable
eigenvectors, for eigenvectors of $\hat{{\bf S}}$ we must
use delta function normalization 
 on $\mathbb{S}^2$ (cf Eq.~(\ref{3.9})).
 
 Based on the actions given in Eqs.~(\ref{3.1}),(\ref{3.2}), we can easily construct the
corresponding (ideal) eigenvectors in ${\cal H}_1$. To 
 handle $\hat{W}$, we need to choose, for each $\hat{{\bf
k}}\in \mathbb{S}^2$, a pair of transverse mutually
orthogonal circular polarization 
 vectors $\mbox{\boldmath$\epsilon$}^{(\pm)}(\hat{{\bf
k}})$. In terms of the spherical polar angles $\theta,
\varphi$ of $\hat{{\bf k}}\in 
 \mathbb{S}^2$, their definitions and important properties
are as follows ( with $C$ for $\cos$ and $S$ for $\sin$):
  \begin{align}
&\mbox{\boldmath$\epsilon$}^{(+)}(\hat{{\bf k}})=\dfrac{e^{i\varphi}}{\sqrt{2}}(C
\theta C\varphi -iS\varphi, C
\theta S\varphi+iC\varphi,
-S\theta),\nonumber\\
&\mbox{\boldmath$\epsilon$}^{(-)}(\hat{{\bf
k}})=i\mbox{\boldmath$\epsilon$}^{(+)}(\hat{{\bf k}})^\ast
\nonumber\\ &~~~~~~~~~~= i\dfrac{e^{-i\varphi}}{\sqrt{2}}(C
\theta C\varphi +iS\varphi, C
\theta S\varphi-iC\varphi,
-S\theta);\nonumber\\
& \hat{{\bf k}}\cdot \mbox{\boldmath$\epsilon$}^{(a)}(\hat{{\bf k}})=0,~a=\pm ~ ; 
~~\mbox{\boldmath$\epsilon$}^{(a)}(\hat{{\bf k}})^\ast \cdot 
\mbox{\boldmath$\epsilon$}^{(b)}(\hat{{\bf k}})=\delta_{a,b};\nonumber\\
& \hat{{\bf k}}\wedge
\mbox{\boldmath$\epsilon$}^{(a)}(\hat{{\bf
k}})=-ia\mbox{\boldmath$\epsilon$}^{(a)}(\hat{{\bf k}})
; ~~\mbox{\boldmath$\epsilon$}^{(+)}(\hat{{\bf k}}) \wedge 
\mbox{\boldmath$\epsilon$}^{(-)}(\hat{{\bf k}})=\hat{{\bf k}}.\nonumber\\
\end{align}
As is well known, transverse circular polarization vectors
defined smoothly all over $\mathbb{S}^2$ do not exist~\cite{
nityananda-ap-14, mukunda-josaa-14, arvind-pla-17}.
The
above choices are well 
defined at $\theta=0$ but multivalued at $\theta=\pi$. Their
behaviours under parity are useful, and read:
\begin{eqnarray}
 \mbox{\boldmath$\epsilon$}^{(a)}(-\hat{{\bf
k}})&=&iae^{2ia\varphi}\mbox{\boldmath$\epsilon$}^{(-a)}(\hat{{\bf
k}}), a=\pm ~,
\end{eqnarray}
so
\begin{eqnarray}
 \hat{{\bf k}}\wedge
\mbox{\boldmath$\epsilon$}^{(a)}(-\hat{{\bf
k}})&=&ia\mbox{\boldmath$\epsilon$}^{(a)}(-\hat{{\bf k}}).
\end{eqnarray}
After some straightforward analysis, the (ideal)
simultaneous eigenvectors of  $\hat{S}_j,\hat{W}$  can be
found upto arbitrary `radial' functions:
\begin{eqnarray}
&{\bf s} \in \mathbb{S}^2,\, w=\pm 1 \longrightarrow |{\bf
s}, w\rangle \in {\cal H}_1:\nonumber\\
&\hat{{\bf S}}|{\bf s},w \rangle = \hbar {\bf s}|{\bf
s},w \rangle,\quad {\hat W}|{\bf s},w \rangle =\hbar w |{\bf s},w
\rangle; \nonumber\\
&(|{\bf s},w \rangle)_j({\bf k})= a(k,{\bf s},w)
\delta^{(2)}(\hat{{\bf k}}, w{\bf s})
\mbox{\boldmath$\epsilon$}^{(+)}_j({\bf s}),~~
 \text{any}~~ a(k,{\bf s},w).\nonumber\\
\label{4.5}
 \end{eqnarray}
 The inner products have the form expected from orthonormality:
 \begin{eqnarray}
 \langle {\bf s}^\prime,w^\prime |{\bf s},w \rangle &=& \int
d^3k (|{\bf s}^\prime,w^\prime \rangle)_j({\bf k})^\ast
(|{\bf s},w \rangle)_j({\bf k})
 \nonumber\\
 &=& \delta_{w,w'} \delta^{(2)}({\bf s}^\prime,{\bf s})
\int_0^\infty k^2 dk a^\prime (k,{\bf s},w)^\ast a (k,{\bf
s},w).\nonumber\\
 \label{4.6}
 \end{eqnarray}

As for the completeness property, we omit the factor $
a(k,{\bf s},w)$ in Eq.~(\ref{4.5}) and find for the angular
part: 
\begin{eqnarray}
 &&\sum_{w=\pm 1} \int_{\mathbb{S}^2} d\Omega({\bf s})
\left(\delta^{(2)}(\hat{{\bf k}}, w{\bf s})
\mbox{\boldmath$\epsilon$}^{(+)}_j({\bf s})\right)
\left(\delta^{(2)}(\hat{{\bf k}^\prime}, w{\bf s})
\mbox{\boldmath$\epsilon$}^{(+)}_{j^\prime}({\bf
s})\right)^\ast \nonumber\\
&&\quad\quad\quad= \delta^{(2)}(\hat{{\bf k}}, {\bf
k}^\prime)\left(\delta_{jj^\prime} -\dfrac{k_j
k_{j^\prime}}{|{\bf k}|^2}\right)
\end{eqnarray}

This is to be compared to the last line in~(\ref{3.9})~:
while the right hand sides are the same, the left hand sides
have very different structures, 
due to the differences between $\hat{{\bf J}}$ and $\hat{{\bf S}}$. 

The fact that $\hat{{\bf S}}$ has continuous eigenvalues
(not at all like a quantum mechanical angular momentum),
hence no normalisable eigenvectors, 
has important consequences. We illustrate this by examining
the properties of $\hat{{\bf S}}$ in a normalized
simultaneous eigenvector 
of $\hat{J}_3$ and $\hat{W}$. This has the general form : 
\begin{align}
 & \hat{J}_3\rightarrow \hbar m, ~\hat{W} \rightarrow \hbar w : \nonumber\\
 & {\bf v}_{m,w}({\bf k}) = a(k,m, w, \theta)e^{i(m-w)\varphi}
 \mbox{\boldmath$\epsilon$}^{(w)}(\hat{{\bf k}}); \nonumber\\
 &\langle {\bf v}_{m,w}|{\bf v}_{m,w}\rangle  \nonumber\\ 
 &~~~=2\pi \int_0^{\infty} k^2 dk \int_0^{\pi} \sin\theta
d\theta |a(k,m,w,\theta)|^2=1.
\label{4.8}
 \end{align}
Here $a(k,m,w,\theta)$ is arbitrary. Let us now define an
associated probability distribution $p(x)$ over $[-1,1]$ in
the polar angle 
$\theta$ with $x=\cos\theta$, as follows:
\begin{align}
 p(x) &=2\pi \int_0^\infty k^2dk |a(k,m,w,\theta)|^2 \geq 0,\nonumber\\
& \int_{-1}^{1}dx ~p(x) =1 . 
\end{align}
The normalisation condition~(\ref{4.8}) implies that
$p(x)$ is not of delta function type, so it describes a non
trivial spread and variance in $x$. 
Then using Eqs.~(\ref{3.2}) and (\ref{4.8}) we find the expectation values of the SAM:
\begin{eqnarray}
 \langle {\bf v}_{m,w}|\hat{S}_l|{\bf v}_{m,w}\rangle
&=&\hbar \int_0^\infty k^2dk \int
_{\mathbb{S}^2}d\Omega({\bf k}) |a(k,m,w,\theta)|^2
\hat{k}_l
 \nonumber\\
 &=&\hbar \langle x\rangle_0\delta_{l,3}, \nonumber\\
 \langle f(x)\rangle_0 &= &\int_{-1}^{1} dx f(x)p(x).
 \label{4.10}
\end{eqnarray}
Going a step further, we can obtain the expectation values
of quadratics in the `spin' components as a $3\times 3$
matrix :
\begin{align}
 &\langle {\bf v}_{m,w}|\hat{S}_l\hat{S}_n|{\bf v}_{m,w}\rangle\nonumber\\
 & =\hbar ^2 \left( \int_0^\infty k^2dk \int
_{\mathbb{S}^2}d\Omega({\bf k}) |a(k,m,w,\theta)|^2
\hat{k}_l\hat{k}_n\right)\nonumber\\
 &= \hbar^2  \text{diag} \left( \frac{1}{2}
\langle(1-x^2)\rangle_0, \frac{1}{2}
\langle(1-x^2)\rangle_0,  \langle x^2\rangle_0\right)
\end{align}
Therefore the SAM variance matrix in the normalized state
$|{\bf v}_{m,w}\rangle$ is, using~(\ref{4.10}), 
\begin{eqnarray}
 V&=&  \hbar^2 \text{diag} \left( \frac{1}{2}
\langle(1-x^2)\rangle_0, \frac{1}{2}
\langle(1-x^2)\rangle_0,  \langle (\Delta
x)^2\rangle_0\right),
 \nonumber\\
 &&(\Delta x)^2= \langle x^2\rangle_0- \langle x\rangle_0^2.
\end{eqnarray}
From the statements made above regarding the nature of the
probability distribution $p(x)$, it is clear that the spread
$(\Delta x)^2$ in 
$\hat{S}_3$ is strictly positive, $(\Delta x)^2 >0$. So in
any normalized state $|{\bf v}_{m,w}\rangle$ with well
defined $\hat{J}_3$ and 
$\hat{W}$, there is always a spread in the values of the
components of $\hat{S}$. In particular even though both
$\hat{J}_3$ and 
$\hat{W}$  commute with $\hat{S}_3$, the normalised
eigenvector $|{\bf v}_{m,w}\rangle$ of $\hat{J}_3$ and
$\hat{W}$ can never be a simultaneous 
eigenvector of $\hat{S}_3$ as well, whatever be the choice
of $a(k,m,w,\theta)$. By the same token, the state $|{\bf
v}_{m,w}\rangle$  
can never be an eigenvector of the third component
$\hat{L}_3$ of OAM, for any choice of $a(k,m,w,\theta)$. 
\section{Paraxial regime and vector Laguerre-Gauss modes}
\label{paraxial}
In the previous Sections we have discussed on the one hand
the exact simultaneous eigenfunctions of the total squared
angular momentum 
 $\hat{{\bf J}}^2$ and its component $\hat{J}_3$, and on the
other hand those of the three components of the SAM
$\hat{{\bf S}}$ and the 
 helicity $\hat{W}$. These 
 are collected together in Eqs.~(\ref{3.9}) and
Eqs.~(\ref{4.5}), (\ref{4.6}) respectively. In both cases,
only
angular and polarization dependences 
 are involved. In the general $\hat{{\bf S}}, \hat{W}$
eigenfunction in Eq.~(\ref{4.5}) for example, an arbitrary,
`radial' function
 $a(k,s,w)$ appears. Similarly in the general simultaneous
eigenvector of  $\hat{J}_3,\hat{W}$  in Eq.~(\ref{4.8}) an
arbitrary function
 $a(k,m,w,\theta)$ is present. 
 
 Now we turn to the physically very important paraxial
regime. As argued in earlier work \cite{oam1}, it is reasonable to
consider the paraxial 
 limit after canonical quantization has been completed and
the photon picture of light has been obtained. Thus once
Eqs.~(\ref{CCR}) and their 
 consequences and interpretation are in hand, in the
subsequent analysis based on Eqs.~(\ref{2.14}) we limit the
choices of ${\bf v}({\bf k})\in 
 {\cal M}$ to those 
 having the paraxial property. That is, the paraxial
approximation is made on the choice of ${\bf v}({\bf k})$
within $\hat{a}({\bf v})$ and 
 $\hat{a}({\bf v})^\dagger$, not in the canonical
quantization rule ${\bf v}({\bf k})\rightarrow
\sqrt{\hbar}\hat{{\bf a}}({\bf k}), 
 {\bf v}({\bf k})^\ast\rightarrow \sqrt{\hbar}\hat{{\bf
a}}({\bf k})^\dagger$ in any sense. 'Paraxial photons' are
to be understood in this way. 
 
 The paraxial region in wave vector space is defined ( in an
approximate way) as consisting of those ${\bf k}$ vectors
whose transverse components 
 ${\bf k}_{\perp}$ are much smaller than their (positive)
longitudinal components : 
 \begin{equation}
  |{\bf k}_{\perp}| << k, ~~k_3\simeq k- k_{\perp}^2/2k. 
 \end{equation}
 A photon wave function ${\bf v}({\bf k})$ is paraxial if it
is negligible outside the paraxial region:
 \begin{equation}
  {\bf v}({\bf k})\simeq 0~\text{unless}~{\bf k}~\text{paraxial}.
 \end{equation}
In that case, transversality determines $v_3({\bf k})$ in
terms of $v_\perp({\bf k})$:
\begin{equation}
 v_3({\bf k}_\perp, k)  \simeq -\left(1+
\frac{k_\perp^2}{2k^2}\right)\frac{{\bf k}_\perp\cdot {\bf
v}_\perp({\bf k}_\perp,k)}{k}.
\end{equation}
The longitudinal component is one order of magnitude smaller
than the transverse components. 

One way in which the paraxial property for ${\bf v}({\bf
k})$ can be achieved is if each component $v_j({\bf k})$ is
a common transverse 
Gaussian  factor times a polynomial in ${\bf k}_\perp$. This
requires that there be a transverse width $w_0$ and some
minimum wave  vector 
magnitude $k_{\text{min}}>0$., and 
\begin{align}
&{\bf v}({\bf k}_\perp,k) =\begin{pmatrix} {\bf a}_\perp
({\bf k}_\perp,k)\cr c({\bf k}_\perp,k)\end{pmatrix}
e^{-w_0^2k_\perp^2/4},\nonumber\\
&~~~w_0 >>\lambda_{\text{max}}=2\pi/k_{\text{min}},\nonumber\\
&c({\bf k}_\perp,k) \simeq -\left(1+
\frac{k_\perp^2}{2k^2}\right)\frac{{\bf k}_\perp\cdot {\bf
a}_\perp({\bf k}_\perp,k)}{k},
 \label{5.4}
\end{align}
with ${\bf a}_\perp$ and $c$ polynomial in ${\bf k}_\perp$. 

We can now connect with the exact $\hat{J}_3$-$\hat{W}$
eigenfunctions in Eq.~(\ref{4.8}), and their paraxial
limits, in this way. For 
given eigenvalues, $\hbar m$, $\hbar w$ of $\hat{J}_3$,
$\hat{W}$ the eigenfunction in Eq.~(\ref{4.8}) contains the
arbitrary function 
$a(k,m,w,\theta)$ as a factor. To make this eigenfunction
paraxial means to impose suitable conditions on this free
function. The paraxial 
( small $\theta$ ) limits of
$\mbox{\boldmath$\epsilon$}^{(\pm)}(\hat{{\bf k}})$ are : 
\begin{eqnarray}
 &&\mbox{\boldmath$\epsilon$}^{(+)}(\hat{{\bf k}})\simeq \frac{1}{\sqrt{2}}
 \begin{pmatrix} 1\cr i\cr-\theta e^{i\varphi}\end{pmatrix}; ~~ 
  \mbox{\boldmath$\epsilon$}^{(-)}(\hat{{\bf k}})\simeq
 \frac{1}{\sqrt{2}}
 \begin{pmatrix} i\cr 1\cr -i\theta e^{-i\varphi}\end{pmatrix}.\nonumber\\
 \label{5.5}
\end{eqnarray} 
In scalar paraxial optics the important family of
Laguerre--Gaussian (LG) mode functions have the general
structure of~(\ref{5.4})--polynomials times 
a Gaussian factor in transverse variables. These are defined
using cylindrical coordinates, so we have the connection :
\begin{eqnarray}
 &&{\bf k} =k (\sin\theta \cos \varphi, \sin\theta \sin
\varphi, \cos\theta)=(\rho\cos\varphi,\rho\sin\varphi,
k_3):\nonumber\\
 && \rho = k\sin\theta,~ k_3=k\cos\theta, ~k_\perp^2=\rho^2, ~k^2= \rho^2+k_3^2. 
 \end{eqnarray}
 For small $\theta$, we have $\rho\simeq k\theta$, $k_3
\simeq k -\rho^2/2k$. The LG mode functions are labelled by
two integers: 
 $p=0,1,2,\cdots, m=0,\pm 1, \pm 2\cdots$; and they are 
\begin{align}
 &\phi_{m,p}({\bf k_\perp})=\dfrac{w_0}{\sqrt{2\pi}}\sqrt{\dfrac{p!}{(p+|m|)!}}~
 ~e^{im\varphi}\left(\dfrac{iw_0\rho}{\sqrt{2}}\right)^{|m|}
\nonumber\\ &~~~~~~~~~~~~~~\times
 L_p^{|m|}\left(\dfrac{w_0^2\rho^2}{2}
 \right)e^{-w_0^2\rho^2/4}.
 \label{5.7}
\end{align}
Comparing Eq.~(\ref{5.4}) with
Eqs.~(\ref{4.8}),(\ref{5.5}),(\ref{5.7}) we are led for
each given $m$ to two choices :  \begin{eqnarray}
\underline{w=1}&&~~~~~~~~~~~~~~a(k,m,+1,\theta)\rightarrow
\phi_{m-1,p}({\bf k_\perp}) :\nonumber\\ && {\bf
v}_{m,+1,p}({\bf k}_\perp,k) = \frac{1}{\sqrt{2}}
\begin{pmatrix} 1\cr i\cr-\theta
e^{i\varphi}\end{pmatrix}\phi_{m-1,p}({\bf
k}_\perp);~~~(a)\nonumber\\
\underline{w=-1}&&~~~~~~~~~~~~~~a(k,m,-1,\theta)\rightarrow
\phi_{m+1,p}({\bf k_\perp}) :\nonumber\\ && {\bf
v}_{m,-1,p}({\bf k}_\perp,k) = \frac{1}{\sqrt{2}}
\begin{pmatrix} i\cr 1\cr-i\theta
e^{-i\varphi}\end{pmatrix}\phi_{m+1, p}({\bf
k}_\perp).~(b)\nonumber\\ \end{eqnarray} To leading
paraxial order, these are the complete--transverse vector LG
mode fields. We stress that these are eigenfunctions of the
total angular momentum component $\hat{J}_3$ and helicity
$\hat{W}$ with respective eigenvalues $\hbar m, \pm \hbar$.
In addition to the labels $m,p$ in Eq.~(\ref{5.7}) in the
scalar case, now the third helicity label $w=\pm $ also
appears.
\section{Concluding remarks}
\label{conc}
We have presented a careful analysis of the
properties of the so-called spin and orbital angular momenta
of light, in the quantum domain, as they apply to single
photon states. It has been known for some time that these
operators, which are hermitian constants of motion, do not
have the spectral properties expected of an angular momentum
in the sense of quantum mechanics. Thus the photon spin is
not such an angular momentum. Its components do not have
discrete quantised eigenvalues. It is a result of
transversality of the Maxwell field that there is no
position operator for the photon, therefore no way of
separating the total angular momentum into well defined and
independent spin and orbital parts. The terms 'spin' and '
orbital' angular momenta of light  are  thus  misnomers
which however cannot now be corrected.

  We show by explicit construction that there exist
ideal (non normalisable) eigenvectors for all three spin
components simultaneously. One can, of course, construct
normalisable wave packets out of these eigenvectors,
involving small patches over the sphere $S^2$. At the
classical level it is an interesting challenge to produce
wave fields corresponding to such solutions of the Maxwell
equations. The helicity and the three spin components do
possess simultaneous ideal eigenvectors, with their
eigenvalues being chosen independently. However a normalised
eigenvector of a component of the total angular momentum and
helicity can never be an eigenvector of that component of
the spin as well.
 
 We recall that a noteworthy feature of the formalism
developed in~\cite{oam1} and briefly recapitulated here is the
one to one correspondence between classical radiation field
configurations and the quantum description thereof at the
single photon level. This leads one to expect that some of
the results arising from the peculiar features of the `spin'
and `orbital' angular momentum operators at the one photon
level, as discussed here ought to have measurable signatures
at the classical level as well.  `

Finally we draw attention to the paraxial vectorial
Laguerre-Gauss fields which are a physically relevant and
nontrivial generalisation of the enormously useful scalar
paraxial mode fields of the same name. It is an experimental
challenge to create such fields, and to bring out their
characteristic signatures.
\section{Acknowledgements} NM thanks the Indian National
Science Academy for the INSA Distinguished Professorship,
during the tenure of which this work was initiated. 
Arvind acknowledges the financial
support from DST/ICPS/QuST/Theme-1/2019/General Project
number {\sf Q-68}.
%

\end{document}